\documentclass{emulateapj} 
\newcommand{\bl}[1]{\mbox{\boldmath$ #1 $}}

\slugcomment{Accepted by The Astrophysical Journal Letters}
\shorttitle{Formation of giant planets}
\shortauthors{Vorobyov \& Basu} 

\begin{document}

\title{Formation and survivability of giant planets on wide orbits}
\author{Eduard I. Vorobyov\altaffilmark{1,}\altaffilmark{2} and Shantanu Basu\altaffilmark{3}}
\altaffiltext{1}{Institute for Computational Astrophysics, Saint Mary's University,
Halifax, NS, B3H 3C3, Canada; vorobyov@ap.smu.ca.} 
\altaffiltext{2}{Research Institute of Physics, Southern Federal University, Stachki 194, Rostov-on-Don, 
344090, Russia.} 
\altaffiltext{3}{Department of Physics and Astronomy, The University of Western Ontario, London, ON, N6A 3K7, Canada; basu@astro.uwo.ca.}


\begin{abstract}
Motivated by the recent discovery of massive planets on wide orbits, we present
a mechanism for the formation of such planets via disk fragmentation in the embedded
phase of star formation. In this phase, 
the forming disk intensively accretes matter from the natal cloud core and 
undergoes several fragmentation episodes.
However, most fragments are either destroyed or driven into the innermost regions 
(and probably onto the star) due to angular momentum exchange with spiral arms, leading to
multiple FU-Ori-like bursts and disk expansion. Fragments that are 
sufficiently massive and form in the late embedded phase (when the disk conditions are less extreme) may open
a gap and evolve into giant planets on typical orbits of several tens to several hundreds 
of AU.
For this mechanism to work, the natal cloud core must have sufficient mass 
and angular momentum 
to trigger the burst mode and also form extended disks of the order of several hundreds of AU. 
When mass 
loading from the natal cloud core diminishes and the main fragmentation phase ends, such extended 
disks undergo a transient episode of contraction and density increase, during which they may give 
birth to a last and survivable set of giant planets on wide and relatively stable orbits.
\end{abstract}


\section{Introduction}
The likelihood of giant planet formation via direct gravitational instability of circumstellar 
disks around solar-type stars has been the subject of intense research in the past years. 
Despite much effort in this field, increasingly sophisticated numerical hydrodynamics simulations
and analytical considerations continue to yield conflicting results. On one hand, some studies indicate
that giant planets can form in massive disks, particularly in their outer parts where conditions
for disk fragmentation are less extreme and the competing core-accretion model is less viable 
\citep[e.g.,][]{Johnson03,Stamatellos07,Mayer07,Boss08,Dodson09,Nero09}.
On the other hand, many studies show that gravitational fragmentation is unlikely, particularly in the
inner few tens of AU due to insufficient disk cooling and strong stellar/envelope irradiation 
\citep[e.g.,][]{Matzner05,Rafikov05,Boley06,Boley07,Rafikov07,Stamatellos08,Cai08}

In spite of a great deal of sophistication, the aforementioned studies miss one important 
aspect---circumstellar disks are {\it not} isolated in the early embedded phase of star formation (hereafter,
EPSF). In this stage,
they are subject to intense mass loading from a natal cloud core, which can significantly alter
the disk's ability to fragment. A self-consistent handling of this process in numerical simulations
is not easy and requires a much larger spatial scale (than just that of the disk).  A spatial resolution
of less than 1~AU is usually needed for planetary-mass fragments to form and {\it survive}. 
Global numerical hydrodynamics simulations of the gravitational
collapse and fragmentation of molecular clouds have demonstrated that forming stars are indeed 
surrounded by accreting
gravitationally unstable disks \citep[e.g.,][]{Bate03,Krumholz07}. Yet, these simulations resolve 
only massive disks, which, if fragmented, produce brown dwarfs or low-mass stellar companions rather
than giant planets. Moreover, such numerical simulations are very computationally intensive
and are unable to explore a wide parameter space and long evolution times.

On the other hand, semi-analytic models and simplified numerical simulations of the gravitational 
collapse 
of dense cloud cores can explore a wide range of initial conditions and can give us a valuable 
insight into the required conditions for disk fragmentation. Using the thin-disk approximation, we 
were able to self-consistently follow the process of cloud core collapse and star/disk formation 
for at least several Myr after the formation of a central stellar object \citep{VB05,VB06,VB09a}.  
These studies have shown that circumstellar disks may be gravitationally unstable and susceptible 
to fragmentation if the rate of gas deposition
onto the disk from the cloud core is greater than that from the disk onto the star, 
disk viscosity is not too high, and the natal cloud cores 
are characterized by sufficiently large rotation rates. 
More sophisticated numerical hydrodynamics simulations, though with an approximate treatment
of gas infall onto the disk, and semi-analytic studies have confirmed the susceptibility of non-isolated
disks to fragmentation, particularly at large radii \citep[e.g.,][]{Kratter08,Rice09,Boley09a,Boley09b,Clarke09,Rafikov09}.

The feasibility of disk fragmentation and giant planet formation is only one part of the problem. 
The other part is the likelihood
of survival of giant planets formed via disk fragmentation. 
Rapid radial migration due to gravitational 
interaction of a giant planet with a natal gas disk \citep{Goldreich80} 
has traditionally been one of the stumbling blocks for the theory of giant planet formation
and many mechanisms have been proposed to stop this migration in the late evolution phase \citep[see
e.g.,][]{Thommes06,Crida07,Ida08,Matsumura09}. In the early EPSF, this problem may be 
even more severe due to the fact that
disks are more massive and profoundly non-axisymmetric. Indeed, our previous numerical studies
have shown that fragments forming in the EPSF are quickly driven into the inner regions and
probably onto the protostar due to exchange of angular momentum with spiral arms \citep{VB05,VB06}.
We have speculated that only those fragments that form in the late EPSF, 
when gravitational instability starts to gradually decline with time,  may have a chance 
to survive. 

In this paper, we present confirmation that the fragments formed in the EPSF can 
survive through this extreme phase and form giant protoplanets (hereafter, GPPs)
on large, relatively stable orbits. 
This finding is made possible by the employment of expanded computational 
resources, by improvements in the numerical model, and by the use of a wider parameter space 
in comparison to previous works.

\section{Model equations and initial conditions}
Our numerical model is explained in detail in \citet{VB06} and is
briefly summarized below. 
We make use of the thin-disk approximation to compute the gravitational collapse of rotating, 
gravitationally unstable cloud cores. 
We start our numerical integration in the pre-stellar phase, which is 
characterized by a collapsing {\it starless} cloud core with a typical radius of $10^4$~AU, 
continue into the EPSF, which sees the formation of a star/disk/envelope
system, and terminate our simulations in the late T Tauri phase.
The mass accretion rate onto 
the disk, $\dot{M}_{\rm env}$, is not a free parameter but is self-consistently determined 
by the dynamics of the gas in the envelope. 
We introduce a ``sink cell'' at $r_{\rm sc}=8$~AU and impose a free inflow inner boundary condition.
Ninety per cent of the gas that crosses the inner boundary is assumed to land onto 
the central star plus the inner axisymmetric disk at $r<8$~AU. 
The other 10\% of the accreted gas is assumed to be carried away with protostellar jets. 

The basic equations of mass and momentum transport in the thin-disk approximation are
\begin{eqnarray}
\label{cont}
 \frac{{\partial \Sigma }}{{\partial t}} & = & - \nabla_p  \cdot \left( \Sigma \bl{v}_p 
\right), \\ 
\label{mom}
 \Sigma \frac{d \bl{v}_p }{d t}  & = &  - \nabla_p {\cal P}  + \Sigma \, \bl{g}_p 
 + (\nabla \cdot \mathbf{\Pi})_p \, ,
\end{eqnarray}
where $\Sigma$ is the mass surface density, ${\cal P}=\int^{Z}_{-Z} P dz$ is the vertically integrated
gas pressure , $Z$ is the vertical scale height,
$\bl{v}_p=v_r \hat{\bl r}+ v_\phi \hat{\bl \phi}$ is the velocity in the
disk plane, $\bl{g}_p=g_r \hat{\bl r} +g_\phi \hat{\bl \phi}$ is the gravitational acceleration 
in the disk plane, and $\nabla_p=\hat{\bl r} \partial / \partial r + \hat{\bl \phi} r^{-1} 
\partial / \partial \phi $ is the gradient along the planar coordinates of the disk. 
The gravitational acceleration $\bl{g}_p$ includes the gravity of a central point object 
(when formed), the gravity of the inner inactive disk ($r<8$~AU),
and the self-gravity of a circumstellar disk and envelope. 
For the kinematic viscosity $\nu$ that enters the viscous stress tensor $\mathbf{\Pi}$, we
use the usual $\alpha$-prescription, with a spatially and temporally uniform 
$\alpha=0.005$. The latter choice is based on our recent work \citep{VB09a}, 
i.e., $\alpha$ is small enough to not eliminate the burst mode and large
enough to drive significant accretion in the late stages of disk evolution.

Equations~(\ref{cont}) and (\ref{mom}) are closed with a barotropic equation
that makes a smooth transition from isothermal to adiabatic evolution at $\Sigma = \Sigma_{\rm cr}$
\citep{VB06}. For the ratio of specific heats we use $\gamma$=1.4. 
The $\gamma$=5/3 case was explored in \citet{VB06}.
The value of $\Sigma_{\rm cr}$ is calculated during the numerical
simulations as $\Sigma_{\rm cr}=m_{\rm H} \mu \, n_{\rm cr}\,2 Z$, where the critical 
volume number density $n_{\rm cr}$ is set to $10^{11}$~cm$^{-3}$ \citep{Larson} 
and the mean molecular weight $\mu= 2.33$. 
The scale height $Z$ is calculated using the assumption of vertical
hydrostatic equilibrium. We note that $Z$ is an increasing function of radius,
which makes $\Sigma_{\rm cr}$ increase with radius as well. In practice, this means
that the inner disk regions are significantly warmer than the outer regions, since the optically 
thick regime in the inner regions is achieved at lower $\Sigma$. 
This in turn impedes the development of gravitational instability and fragmentation 
in the inner disk, in agreement with more 
sophisticated numerical simulations and theoretical predictions
that directly solve for the energy balance equation \citep[see e.g.,][]{Stamatellos08,Boley09a,Boley09b}.
The use of a spatially varying $\Sigma_{\rm cr}$ is an important improvement of the 
numerical model as compared to our previous works.
Equations~(\ref{cont}) and (\ref{mom}) are solved in polar 
coordinates $(r, \phi)$ on a numerical grid with $256 \times 256$ or $512 \times 512$ (depending on
the model) grid zones using the method of finite-differences.  
The radial points are logarithmically spaced. 

Initially, cloud cores have surface densities 
$\Sigma$ and angular velocities $\Omega$ typical for a collapsing, axisymmetric, magnetically
supercritical core \citep{Basu}, with $\Sigma,\Omega \propto r^{-1}$ at large radii.
Cloud cores are initially isothermal and they are
characterized by a specific ratio of the rotational to gravitational energy 
$\beta=E_{\rm rot}/|E_{\rm grav}|$. We have considered the evolution of 82 cloud cores with 
masses $M_{\rm cl}=0.2-3.0~M_\sun$, energy ratios $\beta=0.2-2.2\times10^{-2}$, and initial gas
temperatures $T=10-18$~K. The initial column density is flattened
near the center and achieves an asymptotic large-radius profile $\Sigma(r)
= k\,c_{\rm s}^2/(G r)$, where $k = \sqrt{A}/\pi$. In our previous papers \citep[e.g.][]{Vor09},
we took $A=2$, but here let it vary in the range 
$A$=2--8, so that models can be more gravitationally 
unstable. 

\section{Formation of giant planets}
\subsection{Conditions for fragmentation}
\label{fragment}
The formation of giant planets via disk fragmentation can only take place if 
the the ratio of the local cooling time $t_{\rm c}$ to the local dynamical time $\Omega^{-1}$ 
is smaller than a few \citep[e.g.,][]{Gammie01,Rice03,Mejia05}.
Our preliminary results with disk cooling, viscous and 
shock heating, and stellar irradiation included in the code indicate that this condition
is satisfied in the EPSF, at least in the disk's outer regions \citep{VB10}. 

Another requirement for disk fragmentation is the Toomre criterion 
$Q=c_{\rm s} \Omega/(\pi G \Sigma)<1$, which states that the gas surface density 
$\Sigma$ should be sufficiently high for a disk to fragment. 
Too high a $\Sigma$, however, may prevent 
fragmentation due to increased opacity and cooling time \citep{Nero09}. 
In other words, there exists minimum and maximum values of $\Sigma$ between which 
the instability and fragmentation are expected to occur.
In numerical  simulations that form disks self-consistently (such as our own),
$\Sigma$ naturally increases from low toward higher values 
during the disk formation phase and the disk may pass through the unstable regime.

The critical density for fragmentation can in principle be achieved if the rate of mass deposition 
onto the disk $\dot{M}_{\rm env}$ is greater than the mass flux in the disk $\dot{M}_{\rm d}$
so that $\Sigma$ quickly increases with time \citep{VB07,Vor09,Kratter09,Boley09a}
and this phase of intense infall (i.e., the EPSF) lasts for 
many dynamical times (so that the GPPs have enough time to form).
Simple analytical estimates for a viscous disk  indicate that 
$\dot{M}_{\rm d}/\dot{M}_{\rm env}\approx 3 \alpha/Q$ for $T=30~K$ \citep{Boley09a},
which implies that the outer disk regions are expected to fragment for values of $\alpha\sim
10^{-2}$. Numerical simulations by \citet{Vor09} also show that $\dot{M}_{\rm env}$ 
is on average several times greater than $\dot{M}_{\rm d}$ in the EPSF.

Observations provide conflicting estimates
as to the duration of the embedded phase $\tau_{\rm em}$, ranging from a~few~$\times 10^{4}$~yr 
to a~few~$\times 10^{5}$~yr \citep{Andre94,Evans09}. From simple analytical grounds it 
follows that $\tau_{\rm em}$ should be linearly proportional to the initial cloud core 
mass $M_{\rm cl}$ and inverse proportional to $T^{3/2}$ and $A$. In addition, such effects as
rotation and magnetic fields may steepen the relationship, particularly for cloud 
cores of solar mass and greater.
The numerically obtained values for $\beta=1.3\times 10^{-2}$, $T$=10~K, and $A$=2 range from 
0.03~Myr for $M_{\rm cl}=0.2~M_\sun$ to 0.5~Myr for $M_{\rm cl}=2.1~M_\sun$ \citep{Vor10}.
These values may decrease by as much as 3--4 for $T$=20~K or $A$=8.
Nevertheless, cloud cores in the aforementioned mass range are expected to have $\tau_{\rm em}$ 
that are considerably longer than typical 
orbital times of 360--3200~yr for radial distances of 50--100~AU and stellar masses of 0.1--1~$M_\sun$.

Another important initial parameter that determines the disk propensity to fragment
is the amount of rotation in the natal cloud core. Indeed, cloud cores with greater rotation rates
would make larger disks since the centrifugal radius  
is proportional to the square of the specific angular momentum,
which would increase disk tendency to fragment (recall that fragmentation tends to 
occur at large radii). 
This has been confirmed by numerical and semi-analytic simulations of disk formation 
\citep{VB06,Kratter08,Vor09,Rice09}.

The above analysis indicates that disks formed from sufficiently massive cloud cores 
with sufficiently high angular momenta
are expected to fragment in the embedded phase, particularly in the outer regions. 
This conclusion has been corroborated by numerical simulations that form disks self-consistently 
\citep{VB05,VB06,VB09a,Vor09,Kratter09} or impose some prescribed rate of mass infall 
onto the disk \citep{Boley09a,Boley09b,Rice09}.
It is therefore not so much the feasibility of fragmentation 
as the likelihood of survival of the forming fragments that we focus in the present study.

\subsection{Numerical results}
In agreement with Section~\ref{fragment},
we have found that the disk propensity to fragment increases along the sequence of increasing 
$M_{\rm cl}$ and $\beta$. This tendency is reflected in both the greater numbers and higher masses 
of the fragments, which form via fragmentation of dense spiral arms. The final masses of 
fragments that form in our simulations lie 
in a wide range starting from a few Jupiter masses and ending with low- and 
intermediate-mass brown dwarfs.
For disk fragmentation to take place, the minimum cloud core mass should be 
$M_{\rm cl,min}\approx0.8~M_\sun$ for $\beta=0.2\times10^{-3}$ and $A$=2. For 
$\beta=2\times10^{-2}$ and $A=2$, the corresponding $M_{\rm cl,min}$ is approximately 
0.2~$M_\sun$. These minimum masses decrease by about 30\% for higher perturbation amplitudes $A=8$
due to an associated increase in $\dot{M}_{\rm env}$, which promotes disk fragmentation.
On the other hand, our preliminary results suggest that $M_{\rm cl,min}$ may increase somewhat 
if a more accurate treatment of the thermal physics is considered.

Not all fragments evolve ultimately to GPPs, most have either migrated through the inner
boundary or dispersed in the outer regions (possibly due to insufficient resolution).
Only 6 out of the 82 models have formed GPPs with final masses in the 
$M_{\rm pl}=5-10~M_{\rm J}$ range on relatively stable orbits at 25--200~AU. 
In two models we saw two GPPs forming simultaneously, but in both cases the outer one
had dispersed just after 1.0~Myr of evolution, either due to insufficient numerical resolution or tidal
disruption by the inner GPP because of proximity to the 4:1 resonance.
These numbers should not be treated as representative, since 
we expect the number of survived GPPs to increase in simulations with a higher numerical 
resolution.

Among 47 models with $\beta <10^{-2}$ only two models with high density perturbation 
amplitude $A=8$ (as opposed to standard $A$=2) have revealed planet formation.  
The most likely reason why low-$\beta$ and low-$A$ models fail to form GPPs is that
their disks are too small and the fragments are forming too close to the star, which lowers their
chances to survive. In addition, these models have lower disk-to-star mass ratios 
than their high-$\beta$ and high-$A$ counterparts, which results in the formation 
of lower-mass fragments. Such fragments need more time to open the gap and slow down the
fast inward migration in the embedded phase.

\begin{figure*}
\includegraphics[width=17cm]{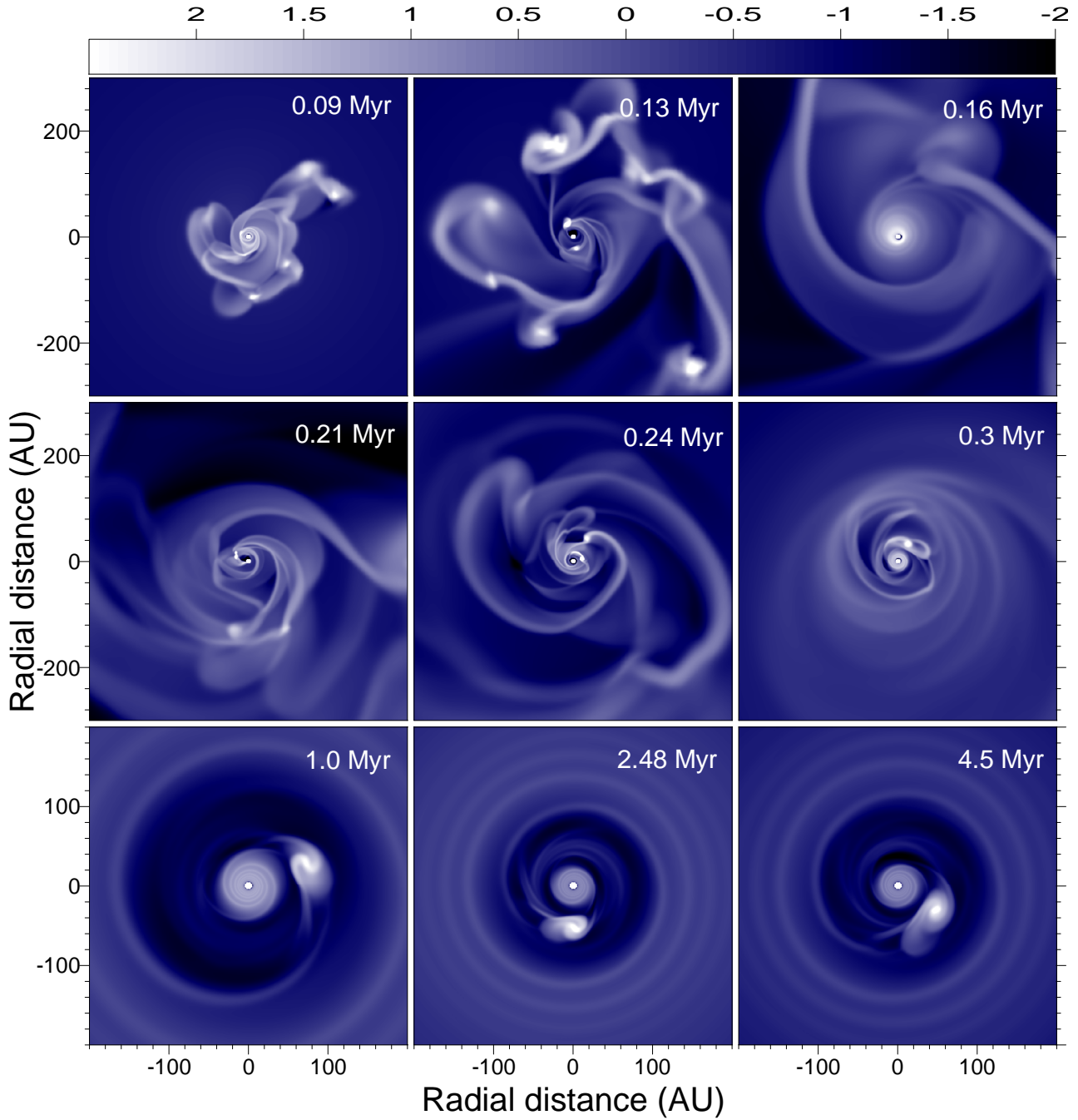}
      \caption{Gas surface density maps (g~cm$^{-2}$, log units) at six times after 
      the formation of the central
      star (bright circle in the coordinate center) in model~1 
      ($M_{\rm cl}=0.9~M_\sun$, $\beta=1.3\times10^{-2}$, $A$=2). 
      Note that we zoom in as the time increases. The top two rows contain images of size 600 AU on
      each side, while the bottom row contains images of size 400 AU on each side.}
         \label{fig1}
\end{figure*}

Figure~\ref{fig1} presents the gas surface density maps (in g~cm$^{-2}$) in  model~1
($M_{\rm cl}=0.9~M_\sun$, $\beta=1.3\times10^{-2}$, $A$=2) at six evolution times after the formation
of the disk. The rotation is counterclockwise (note that we zoom in at the bottom row).
Several fragments condense in the outer parts of the spiral arms as early as 0.09~Myr after 
the disk formation, but none of them have survived by the end of the embedded phase at 
$\approx0.16$~Myr when about 75--80\% of the envelope has been accreted by the disk. 
They are all driven into the sink cell via a very efficient
exchange of angular momentum with the spiral arms, possibly leading to multiple FU Orionis bursts 
\citep{VB05,VB06} or forming giant planets on very close orbits. 
The byproduct of these bursts is disk expansion due to the conservation of angular
momentum. When  the mass loading from the envelope diminishes and the burst phase ends, this
expansion is followed by transient disk contraction, during which gas surface density 
increases and several more fragments form in the disk's
outermost regions ($t=0.21$~Myr). These fragments quickly migrate in the inner 100~AU and, 
by $t=0.3$~Myr, only one fragment survives, which later opens a gap and evolves into a  
well defined GPP possessing its own {\it counterrotating} minidisk. Such counterrotating 
minidisks are seen around many fragments. We believe that this
effect is caused by the gravitational capture of some of 
the neighboring material, which receives a counterrotating twist around the forming fragment 
due to differential rotation of the natal spiral arm.

\begin{figure}
  \resizebox{\hsize}{!}{\includegraphics{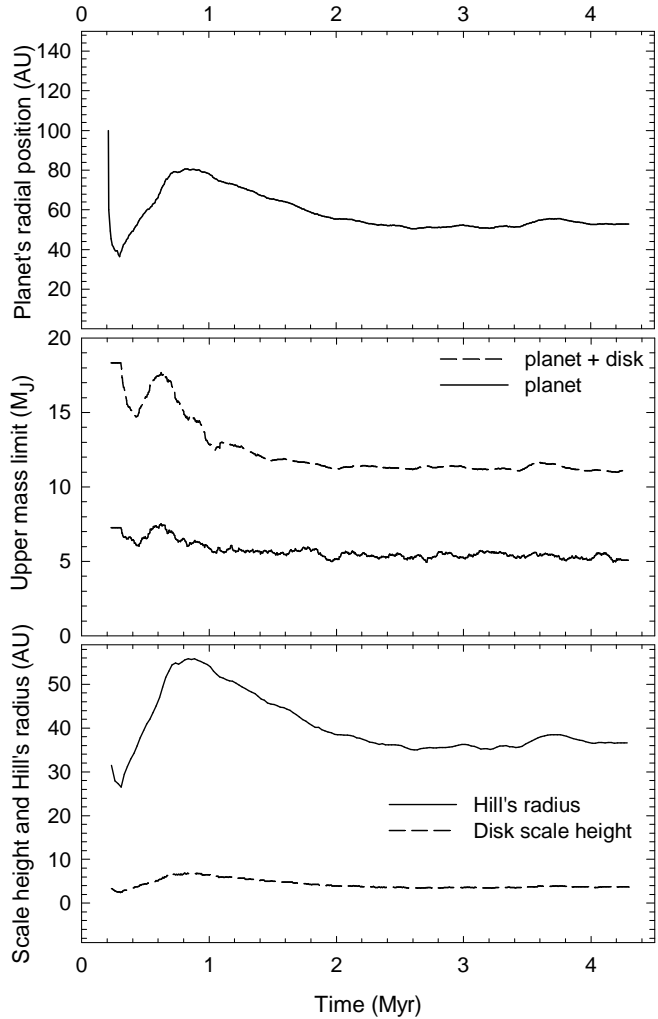}}
      \caption{Planet's radial position (top), mass (middle), and Hill's radius and vertical scale height (bottom), for model 1.}
         \label{fig2}
\end{figure}

Figure~\ref{fig2} shows the GPP's radial position $r_{\rm pl}$ (top), 
mass (middle), and Hill's radius (bottom), as a function of time in model~1. 
Since we do not use sink particles for GPPs, these
values should be treated as approximate. Upon its formation at $r_{\rm p}\ga100$~AU, the GPP migrates quickly in the inner 40~AU and back (this time more slowly) to $r\approx 80$~AU. 
The outward migration is then followed by a gradual inward migration until 
the GPP finally settles at $r_{\rm p}\approx 52$~AU.

The mass of the GPP is estimated by integrating the azimuthally-averaged gas surface 
density profile around a local maximum at the planet location. The middle panel of 
Fig.~\ref{fig2} shows
the {\it upper} limits on the total mass of the inner GPP plus its mini-disk (dashed line) 
and the mass $M_{\rm pl}$ of the inner GPP (solid line). 
The latter value is uncertain to within a factor of unity.
The total (planet plus disk) mass is always in the giant planet regime, while
$M_{\rm pl}$ is around $5\pm 1~M_{\rm J}$. The bottom panel shows the Hill radius 
$r_{\rm H}=r_{\rm pl}(M_{\rm pl}/3(M_\ast+M_{\rm pl}))^{1/3}$ of the GPP (solid line) 
and the azimuthally-averaged vertical scale height at the planet position 
${\overline Z}$ (dashed line). It is seen that
a condition for gap opening, $r_{\rm H} > \overline{Z}$, is satisfied.

\begin{figure}
  \resizebox{\hsize}{!}{\includegraphics{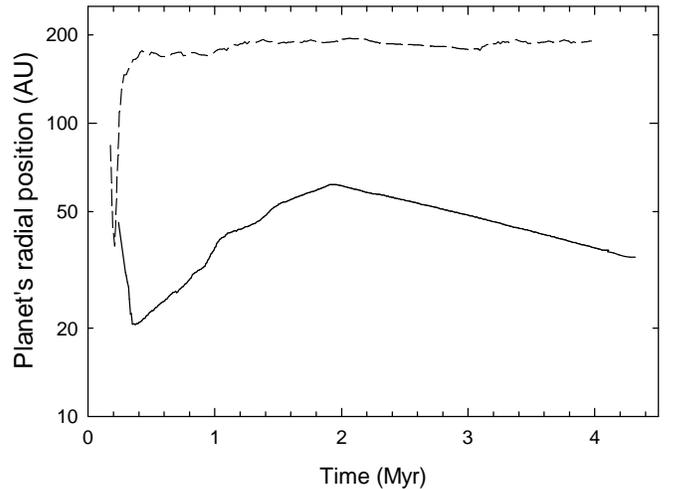}}
      \caption{Planet's radial position in model~2 (solid line, 
      $M_{\rm cl}=0.4~M_\sun$, $\beta=0.01$, $A=2$) and model~3 (dashed line,
      $M_{\rm cl}=2.0~M_\sun$, $\beta=3\times 10^{-3}$, $A=8$).}
         \label{fig3}
\end{figure}

How stable are the orbits of GPPs in our modeling? 
Figure~\ref{fig3} presents the radial position of GPPs vs. time in model~2 (solid line, 
$M_{\rm cl}=0.4~M_\sun$, $\beta=0.01$, $A=2$) and model~3 (dashed line, 
$M_{\rm cl}=2.0~M_\sun$, $\beta=3\times 10^{-3}$, $A=8$). 
%
Initially, model~2 demonstrates a similar migration pattern to that of model~1. However,
the GPP in model~2 does not seem to settle at a stable orbit but slowly migrates inward. 
Whether or not this migration would stop at later times is unclear. On the contrary,
the GPP in model~3 appears to stabilize at a rather large radial distance of 190~AU.
These examples indicate that GPPs may have various migration histories, which depend probably
on particular physical conditions in the disk and the natal cloud core.
The net planet masses are $5\pm1~M_{\rm J}$ in model 2 and $10\pm1~M_{\rm J}$ in model~3.

\section{Conclusions}

We have studied the long-term evolution of disks that are formed by
the self-consistent collapse of prestellar cores. Our model
yields gas giant formation starting from 
initial conditions of the early stages of star formation. The initial cores 
are more gravitationally unstable and have greater angular momenta
than similar models studied in the past, and a large number of 
models have been run with relatively high resolution. 
An early burst mode of evolution is characterized by the formation of
clumps which are then driven into the inner disk. However, in a  
small subset of models, massive fragments are formed on wide
orbits and settle into stable orbits of radius $\gtrsim 50$ AU. 
An interesting feature is that minidisks around these fragments can be 
{\it counterrotating} with respect to the disk.
Sometimes, the final orbit can be much larger or 
alternatively there can be a slowly continuing inward migration. 
We believe that our results can explain the purported observations
of giant planets on wide orbits. By extrapolation, they may also 
represent the first stages of the eventual formation of a low mass 
brown dwarf companion.

\acknowledgments   
   We are thankful to the referee, Dr. Richard Durisen, for an insightful report. 
   EIV gratefully acknowledges present support 
   from an ACEnet Fellowship.  SB was supported by a grant
   from NSERC. Numerical simulations were done 
   on the Atlantic Computational Excellence Network (ACEnet),
   on the Shared Hierarchical Academic Research Computing Network (SHARCNET), and
   at the Center of Collective Supercomputer
   Resources, Taganrog Technological Institute, South Federal University.

\end{document}